\documentclass[11pt,letterpaper]{article}
\usepackage[utf8]{inputenc}
\usepackage[margin=1in]{geometry}
\usepackage{amsmath, amssymb, amsfonts}
\usepackage{graphicx}
\usepackage{amsthm}
\usepackage{enumitem}
\usepackage{bm}
\usepackage{color}
\usepackage{geometry}
\usepackage{complexity}
\usepackage{algorithm}
\usepackage{algpseudocode}
\usepackage{float}
\theoremstyle{plain}

\newtheorem{theorem}{Theorem}[section]
\newtheorem{lemma}[theorem]{Lemma}

\newtheorem{corollary}[theorem]{Corollary}

\theoremstyle{definition}

\makeatletter
\@addtoreset{equation}{section}

\usepackage[hidelinks]{hyperref} 

\makeatother

\title{An Approximation Algorithm for Monotone Submodular Cost Allocation}
\author{Ryuhei Mizutani\thanks{Faculty of Engineering, Tokyo University of Agriculture and Technology, Tokyo, 184-8588, Japan. E-mail: \texttt{mizutani@go.tuat.ac.jp}}}

\date{}

\begin{document}

\maketitle

\begin{abstract}
In this paper, we consider the minimum submodular cost allocation (\textsc{MSCA}) problem.
The input of \textsc{MSCA} consists of $k$ nonnegative submodular functions $f_1,f_2,\ldots,f_k$ on the ground set $N$ given by evaluation oracles, and the goal is to partition $N$ into $k$ (possibly empty) sets $S_1,S_2,\ldots,S_k$ so that $\sum_{i=1}^k f_i(S_i)$ is minimized.
In this paper, we focus on the case when $f_1,f_2,\ldots,f_k$ are monotone, which coincides with the facility location problem with submodular facility costs introduced by Svitkina and Tardos.
We show that the integrality gap of a natural LP-relaxation for \textsc{MSCA} with monotone submodular functions is at most $k/2$, yielding a $k/2$-approximation algorithm.
For fixed $k$, we also provide a matching lower bound on the integrality gap and prove a matching hardness result via an approximation preserving reduction from the minimum vertex cover problem in a $k$-partite hypergraph.
We further provide applications of our results to the dual linear program of weighted $k$-polymatroid intersection and the minimum-weight $b$-vertex cover problem in a $k$-partite hypergraph. 
\end{abstract}

\section{Introduction}
\label{sec:intro}
Let $N$ be a finite set and $k\ge 2$ a positive integer.
A set function $f:2^N\to \mathbb{R}$ is called \textit{submodular} if $f(S)+f(T)\ge f(S\cup T)+f(S\cap T)$ holds for every $S,T\subseteq N$.
In this paper, we focus on the \textit{Minimum Submodular Cost Allocation} (\textsc{MSCA}) problem, introduced by Chekuri and Ene \cite{chekuri2011icalp}.
The input of \textsc{MSCA} is $k$ nonnegative submodular functions $f_1,f_2,\ldots,f_k:2^N\to \mathbb{R}_{\ge 0}$ given by evaluation oracles, and the goal is to  partition $N$ into $k$ (possibly empty) sets $S_1,S_2,\ldots,S_k$ so that $\sum_{i=1}^k f_i(S_i)$ is minimized, i.e.,
\begin{align*}
\min\left\{\sum_{i=1}^k f_i(S_i)\ \middle|\ S_1,S_2,\ldots,S_k\ \mathrm{is\ a\ partition\ of\ }N\right\}.
\end{align*}

Several well-studied problems such as the multiway cut problems in graphs and hypergraphs, the submodular multiway partition problems, and uniform metric labeling can be cast as special cases of \textsc{MSCA} \cite{chekuri2011icalp}.
When all submodular functions are monotone, \textsc{MSCA} is equivalent to the facility location problem with submodular facility costs introduced by Svitkina and Tardos~\cite{svitkina2010}.
In this paper, we focus on this setting, that is, we assume that $f_i(S)\le f_i(T)$ for all $S\subseteq T\subseteq N$ and $i=1,2,\ldots,k$.
We refer to this monotone case of MSCA as \textsc{Monotone-MSCA}.
This problem is NP-hard since the monotone submodular multiway partition problem \cite{Bi2025}, which is known to be NP-hard, is a special case of \textsc{Monotone-MSCA}.
The aim of this paper is to investigate the approximability of \textsc{Monotone-MSCA} and the integrality gap of a natural LP-relaxation when $k$ is fixed.

\paragraph*{LP-relaxation.}
To provide an LP-relaxation for \textsc{Monotone-MSCA}, we introduce some notation.
Let $\mathbf{1}\in \mathbb{Z}^N_{>0}$ denote the all-ones vector and $[k]:=\{1,2,\ldots,k\}$.
For any $S\subseteq N$, let $\chi_S\in \{0,1\}^N$ denote the characteristic vector of $S$.
Throughout this paper, we assume that the input functions $f_1,f_2,\ldots,f_k$ satisfy $f_i(\emptyset)=0$ for every $i\in [k]$.
We consider the following linear programming relaxation for \textsc{MSCA}:
\begin{align*}
(\textsc{LP})\ \ \ \ \ \mathrm{minimize}&\ \ \ \sum_{S\subseteq N,i\in [k]}y_i(S) f_i(S)\\
\mathrm{subject\ to}&\ \ \ y_i\in \mathbb{R}^{2^N}_{\ge 0}\ \mathrm{for\ every\ }i\in [k],\\&\ \ \ \sum_{S\subseteq N,i\in [k]}y_i(S) \chi_S=\mathbf{1}.
\end{align*}
In this LP-relaxation, we interpret $y_i(S)=1$ to mean that $S$ is assigned to $f_i$ in \textsc{MSCA}, whereas $y_i(S)=0$ means that $S$ is not assigned to $f_i$.
We note that if an integer feasible solution $(y_1,y_2,\ldots,y_k)$ satisfies $y_i(S)=y_i(T)=1$ for two distinct sets $S,T\subseteq N$ and $i\in [k]$, then by the submodularity of $f_i$ we can update $y_i(S):=y_i(T):=0$ and $y_i(S\cup T):=1$ without increasing the objective value.
Hence, after such uncrossing operations,  any integer feasible solution to \textsc{(LP)} induces a partition of $N$ for \textsc{MSCA}.

The linear program \textsc{(LP)} is equivalent to the convex program relaxation introduced by Chekuri and Ene \cite{chekuri2011icalp} (see Appendix \ref{app:convex_program}).
Regarding the integrality gap of \textsc{(LP)}, Ene and Vondr\'{a}k~\cite{ene2014} showed that if $\mathrm{P}\ne \mathrm{NP}$, then the integrality gap of \textsc{(LP)} for general submodular functions cannot be upper-bounded by any factor.
Hence, it is natural to focus on special cases such as a monotone case that admit a bounded integrality gap.
From now on, we assume that $f_1,f_2,\ldots,f_k$ in \textsc{$\mathrm{(LP)}$} are monotone submodular, unless otherwise stated.
Svitkina and Tardos \cite{svitkina2010} gave a $(1+\ln n)$-approximation algorithm for \textsc{Monotone-MSCA}, and Chekuri and Ene \cite{chekuri2011icalp} showed that the integrality gap of \textsc{$\mathrm{(LP)}$} is at most $(1+\ln n)$, where $n:=|N|$.
Santiago and Shepherd \cite{santiago2018} showed that the integrality gap of \textsc{$\mathrm{(LP)}$} is at most $k$, yielding a $k$-approximation for \textsc{Monotone-MSCA}.
On the negative side, Svitkina and Tardos \cite{svitkina2010} showed that when $k$ is part of the input, \textsc{Monotone-MSCA} is set-cover hard.
This implies that \textsc{Monotone-MSCA} cannot be approximated within a factor of $(1-o(1))\cdot \ln n$ \cite{feige1998}, which matches the approximation ratio shown by Svitkina and Tardos \cite{svitkina2010}.
While the tight approximation ratio is known when $k$ is part of the input, the approximability and the integrality gap of \textsc{$\mathrm{(LP)}$} for fixed $k$ are much less understood.
In this paper, we aim to obtain the tight  approximation ratio and the integrality gap of \textsc{$\mathrm{(LP)}$} for fixed $k$.

\paragraph*{Our results and techniques.}

As mentioned previously, the current best upper bound on the integrality gap of \textsc{$\mathrm{(LP)}$} is $k$ when $k$ is fixed.
Our result improves this to $k/2$.
\begin{theorem}
\label{theorem:igmost}
Suppose that $f_1,f_2,\ldots f_k$ are monotone submodular functions on $N$ with $f_i(\emptyset)=0$ for each $i\in [k]$.
Then, the integrality gap of \textsc{$\mathrm{(LP)}$} is at most $k/2$.
\end{theorem}

Since the proof of Theorem \ref{theorem:igmost} is constructive, it yields a $k/2$-approximation algorithm for \textsc{Monotone-MSCA}, which improves the current best approximation ratio $k$ for fixed $k$.
\begin{theorem}
\label{thm:approx}
There exists a $k/2$-approximation algorithm for \textsc{Monotone-MSCA}.
\end{theorem}

We also provide a matching lower bound on the integrality gap in Theorem \ref{theorem:igmost}, and prove a hardness result that matches the approximation guarantee in Theorem \ref{thm:approx} under the Unique Games Conjecture.

\begin{theorem}
\label{thm:ig_atleast}
Suppose that $f_1,f_2,\ldots f_k$ are monotone submodular functions on $N$ with $f_i(\emptyset)=0$ for each $i\in [k]$.
Then, the integrality gap of \textsc{$\mathrm{(LP)}$} is at least $k/2-o(1)$.
\end{theorem}

\begin{theorem}
\label{thm:approx_hardness}
Assuming the Unique Games Conjecture, \textsc{Monotone-MSCA} is NP-hard to approximate within a factor of $k/2-\epsilon$ for any $\epsilon >0$ and fixed $k\ge 2$.
\end{theorem}

These results are closely related to the approximability/inapproximability results for the minimum vertex cover problem in a $k$-partite hypergraph.
To define this problem, we introduce some notation.
Let $H=(V,E)$ be a $k$-partite hypergraph with a given $k$-partition $\{V_1,\ldots,V_k\}$ of the vertex set $V$, that is, every edge $e\in E$ satisfies $|e\cap V_i|\le 1$ for all $i\in [k]$.
A subset $X\subseteq V$ is called a \textit{vertex cover} if $e\cap X\ne \emptyset$ for every $e\in E$.
The \textit{minimum vertex cover problem} in a $k$-partite hypergraph $H$ asks to find a vertex cover of the minimum size.
For convenience, we denote this problem as $k$-\textsc{HypVC}.
In this paper, we provide an approximation preserving reduction from $k$-\textsc{HypVC} to \textsc{Monotone-MSCA}.
Via this connection, our approximability results  imply the corresponding results for $k$-\textsc{HypVC}, while our inapproximability results are implications of the corresponding results for $k$-\textsc{HypVC}.

Indeed, Lov\'{a}sz \cite{lovasz1975} showed that the integrality gap of a natural LP-relaxation $(\textsc{LP}_{\text{VC}})$ for $k$-\textsc{HypVC} is at most $k/2$, which yields a $k/2$-approximation algorithm for $k$-\textsc{HypVC}.
Theorems \ref{theorem:igmost} and \ref{thm:approx} imply these results via approximation preserving reductions.
On the hardness side, Aharoni et al.~\cite{aharoni1996} showed that the integrality gap of $(\textsc{LP}_{\text{VC}})$ is at least $k/2-o(1)$ for fixed $k$.
In addition, Guruswami et al. \cite{guruswami2015} showed that assuming the Unique Games Conjecture, $k$-\textsc{HypVC} is NP-hard to approximate within a factor of $k/2-\epsilon$ for any $\epsilon>0$ and fixed $k$.
Theorems \ref{thm:ig_atleast} and \ref{thm:approx_hardness} can be proved by these results via approximation preserving reductions.

To prove Theorem \ref{theorem:igmost}, we consider the integrality gap of a more general linear program \textsc{$(\mathrm{LP}_w)$} obtained by replacing the all-ones vector $\mathbf{1}$ in \textsc{$\mathrm{(LP)}$} by a general integer vector $w$.
We show that the integrality gap of \textsc{$(\mathrm{LP}_w)$} is at most $k/2$, which implies Theorem \ref{theorem:igmost}.
This result is not only useful for proving Theorem \ref{theorem:igmost}, but also has applications to the dual linear program of weighted $k$-polymatroid intersection, and the minimum-weight $b$-vertex cover problem in $k$-partite hypergraphs (see Section~\ref{sec:app}).

To show that the integrality gap of \textsc{$(\mathrm{LP}_w)$} is at most $k/2$, we construct a $k/2$-approximate integer feasible solution to \textsc{$(\mathrm{LP}_w)$}.
Our construction is based on Lov\'{a}sz's proof \cite{lovasz1975} of the upper bound on the integrality gap of $(\textsc{LP}_{\text{VC}})$ as follows.
We first obtain an optimal solution $(y_1,y_2,\ldots,y_k)$ to \textsc{$(\mathrm{LP}_w)$} whose supports are chains via the ellipsoid method.
Then, we partition each chain into several subchains, and take the largest $2/k$ fraction of the sets in each subchain.
Lov\'{a}sz \cite{lovasz1975} constructed several vertex covers from an optimal solution to $(\textsc{LP}_{\text{VC}})$ based on a certain matrix, and one of them is guaranteed to be a $k/2$-approximate  vertex cover.
Similarly, based on the matrix obtained by Lov\'{a}sz \cite{lovasz1975}, we partition the sets chosen above into several groups, so that the sum of the characteristic vectors of the sets in each group is at least $w$.
These groups correspond to integer feasible solutions to \textsc{$(\mathrm{LP}_w)$}, and one of them corresponds to a $k/2$-approximate integer feasible solution.

To prove Theorem \ref{thm:ig_atleast}, we show that the integrality gap of $(\textsc{LP})$ is lower-bounded by that of $(\textsc{LP}_{\text{VC}})$, by appropriately setting $f_1,f_2,\ldots,f_k$.
Combined with the integrality gap lower bound of $(\textsc{LP}_{\text{VC}})$ by Aharoni et al. \cite{aharoni1996}, we obtain the lower bound in Theorem \ref{thm:ig_atleast}.
For the proof of Theorem \ref{thm:approx_hardness}, we show an approximation preserving reduction from $k$-\textsc{HypVC} to \textsc{Monotone-MSCA} by setting $f_1,f_2,\ldots,f_k$ as above.
By this reduction, the approximation hardness result for $k$-\textsc{HypVC} by Guruswami er al. \cite{guruswami2015} implies Theorem \ref{thm:approx_hardness}.

\paragraph*{Related work.}
Svitkina and Tardos \cite{svitkina2010} introduced the facility location problem with submodular facility costs defined as follows.
The input consists of a set of facilities $\mathcal{F}$, a set of demands or clients $\mathcal{D}$, a nonnegative cost $c_{ij}$ to connect facility $i$ to client $j$ for each $i\in \mathcal{F}$ and $j\in \mathcal{D}$, and a facility opening cost function $f_i$ for each $i\in \mathcal{F}$.
The goal is to assign each client to a facility so as to minimize the sum of the facility opening cost and the connection cost, where the facility opening cost for a facility $i$ is a monotone submodular function $f_i$ of the clients assigned to $i$.
Since the sum of the facility opening cost and the connection cost for a facility $i$ is a monotone submodular function of the clients assigned to $i$, the facility location problem defined above is equivalent to \textsc{Monotone-MSCA}.
As mentioned above, Svitkina and Tardos \cite{svitkina2010} gave a $(1+\ln |\mathcal{D}|)$-approximation algorithm for this problem, and showed a matching lower bound $(1-o(1))\cdot \ln |\mathcal{D}|$ via a reduction from the set cover problem \cite{feige1998} when $|\mathcal{F}|$ is part of the input.

Recently, Bi et al. \cite{Bi2025} considered the submodular multiway partition problem for monotone submodular functions, which they call the monotone submodular multiway partition (\textsc{Mono-Sub-MP}).
In \textsc{Mono-Sub-MP}, we are given a nonnegative monotone submodular function $f:2^N\to \mathbb{R}_{\ge 0}$ with $f(\emptyset)=0$ and $k$ terminals $t_1,t_2,\ldots,t_k\in N$. 
The goal is to find a partition $S_1,S_2,\ldots,S_k$ of $N$ with $t_i\in S_i$ for every $i\in [k]$ in order to minimize $\sum_{i=1}^k f(S_i)$.
As the submodular multiway partition problem can be seen as a special case of MSCA, \textsc{Mono-Sub-MP} is a special case of \textsc{Monotone-MSCA} as follows.
Define $N':=N\setminus \{t_1,t_2,\ldots,t_k\}$ and $f_i(S):=f(S\cup \{t_i\})-f(\{t_i\})$ for each $i\in [k]$ and $S\subseteq N'$.
Then, $f_i$ is a nonnegative monotone submodular function with $f_i(\emptyset)=0$ for every $i\in [k]$.
Hence, \textsc{Mono-Sub-MP} can be cast as a special case of \textsc{Monotone-MSCA}.
Bi et al. \cite{Bi2025} showed that \textsc{Mono-Sub-MP} admits a $4/3$-approximation, and does not admit a $(10/9-\epsilon)$-approximation for every constant $\epsilon>0$.
For other recent works related to submodular partition problems, see \cite{abbasi2024,berczi2026,chandrasekaran2023,hirayama2023,santiago2021}.

There are further recent works related to \textsc{MSCA}.
Etesami \cite{etesami2024} showed that multi-item welfare maximization under network externalities can be formulated as a special case of \textsc{MSCA}.
Deng et al.~\cite{deng2023} introduced the generalized load-balancing (\textsc{GLB}) problem, and showed that \textsc{GLB} is equivalent to \textsc{MSCA} when the outer norm is the $\ell_1$-norm.

\paragraph*{Organization.}
The rest of the paper is organized as follows.
In Section \ref{sec:approx}, we show that a generalization of  \textsc{$\mathrm{(LP)}$} has an integrality gap of at most $k/2$, which implies Theorem \ref{theorem:igmost}.
This also yields a $k/2$-approximation algorithm for \textsc{Monotone-MSCA}.
In Section \ref{sec:app}, we provide two applications of our main results.
The first application is an upper bound on the integrality gap of the dual linear program of weighted $k$-polymatroid intersection, and the second application is a $k/2$-approximation algorithm for the minimum-weight $b$-vertex cover problem in a $k$-partite hypergraph.
In Section \ref{sec:app}, we also prove Theorems \ref{thm:ig_atleast} and \ref{thm:approx_hardness} with the aid of the proof in the first application.

\section{Approximation algorithm}
\label{sec:approx}
In this section, we prove Theorems \ref{theorem:igmost} and \ref{thm:approx}.
We show that a generalization of \textsc{$\mathrm{(LP)}$} has an integrality gap of at most $k/2$, which implies Theorem \ref{theorem:igmost}.
This also yields a $k/2$-approximation algorithm for \textsc{Monotone-MSCA}.
The proof is based on that of Lov\'{a}sz \cite{lovasz1975} for the upper bound on the integrality gap of a natural LP-relaxation for $k$-\textsc{HypVC}.

Let $N$ be a finite set of cardinality $n$, and $k\ge 2$ a positive integer.
For each $i\in [k]$, let $f_i:2^N\to \mathbb{R}_{\ge 0}$ be a monotone submodular function with $f_i(\emptyset)=0$.
For an integer vector $w\in \mathbb{Z}_{\ge 0}^{N}$, we define \textsc{$(\mathrm{LP}_w)$} as follows:
\begin{align*}
\textsc{$(\mathrm{LP}_w)$}\ \ \ \ \ \mathrm{minimize}&\ \ \ \sum_{S\subseteq N,i\in [k]}y_i(S) f_i(S)\\
\mathrm{subject\ to}&\ \ \ y_i\in \mathbb{R}^{2^N}_{\ge 0}\ \mathrm{for\ every\ }i\in [k],\\&\ \ \ \sum_{S\subseteq N,i\in [k]}y_i(S) \chi_S=w.
\end{align*}
Note that if $w=\mathbf{1}$, then \textsc{$(\mathrm{LP}_w)$} coincides with \textsc{$\mathrm{(LP)}$}.
The linear program \textsc{$(\mathrm{LP}_w)$} can also be seen as a slight modification of the dual linear program of weighted $k$-polymatroid intersection (see Section \ref{subsec:k-poly} for details).
To prove Theorems \ref{theorem:igmost} and \ref{thm:approx}, we show the following more general results:

\begin{theorem}
\label{thm:ig_most_lattice}
The integrality gap of \textsc{$(\mathrm{LP}_w)$} is at most $k/2$.
\end{theorem}
\begin{theorem}
\label{thm:approx_ex}
There exists a polynomial-time algorithm to find an  integer feasible solution to \textsc{$(\mathrm{LP}_w)$} whose objective value is at most $k/2$ times the optimal value of  \textsc{$(\mathrm{LP}_w)$}.
\end{theorem}

Since \textsc{$(\mathrm{LP}_w)$} includes \textsc{$\mathrm{(LP)}$} as a special case,  Theorems \ref{thm:ig_most_lattice} and \ref{thm:approx_ex} imply Theorems \ref{theorem:igmost} and \ref{thm:approx}, respectively.
We first provide a constructive proof of Theorem \ref{thm:ig_most_lattice}, and then show that the construction can be done in polynomial time, which implies Theorem \ref{thm:approx_ex}.

To prove the upper bound on the integrality gap, we show that a $k/2$-approximate integer solution to \textsc{$(\mathrm{LP}_w)$} can be constructed from an optimal solution to \textsc{$(\mathrm{LP}_w)$}.
Since the coefficient matrix of the constraints in \textsc{$(\mathrm{LP}_w)$} is rational,  \textsc{$(\mathrm{LP}_w)$} has a rational optimal solution $(y_1^*,y_2^*,\ldots,y_k^*)$.
Let $\mathcal{L}_i:=\{S\subseteq N\mid y^*_i(S)>0\}$ and $d_i:=|\mathcal{L}_i|$ for each $i\in [k]$.
By the uncrossing operations, we may assume that $\mathcal{L}_i$ is a chain for each $i\in [k]$.
Let $C_i^1\supsetneq C_i^2\supsetneq \cdots \supsetneq C_i^{d_i}$ be the members of $\mathcal{L}_i$ for each $i\in [k]$.
We note that this is the only part using the submodularity of $f_1,f_2,\ldots,f_k$ in the proofs of Theorems \ref{thm:ig_most_lattice} and \ref{thm:approx_ex}.

For convenience of discussion, we will convert $y^*_i$ into an integer vector $z_i$ by multiplying it by some constant factor, and make $z_i(C_i^\ell)$ copies of $C_i^\ell$  for each $\ell$.
To this end, we define $M$ as the product of $k(k-1)$ and the denominators of $y^*_i(C_i^j)$ for all $i,j$.
Then, $z_i:=M\cdot y^*_i$ is an integer vector for every $i\in [k]$.
Let $b:=\max_{e\in N}w(e)$.
For each $i\in [k]$, we define $U_i^1,U_i^2,\ldots,U_i^{bM}\subseteq N$ by setting $U_i^j:=C_i^1$ for the first $z_i(C_i^1)$ indices $j=1,2,\ldots,z_i(C_i^1)$, and then setting $U_i^j:=C_i^2$ for the next $z_i(C_i^2)$ indices $j=z_i(C_i^1)+1,z_i(C_i^1)+2,\ldots,z_i(C_i^1)+z_i(C_i^2)$, and continue in this way up to $C_i^{d_i}$.
For the remaining indices $j$, setting $U_i^j:=\emptyset$.
In other words, for each $\ell\in [d_i]$, we set $U_i^j:=C_i^\ell$ for all $j$ in the following block of indices:
\begin{align*}
1+\displaystyle\sum_{p=1}^{\ell-1} z_i(C_i^p)\le j\le \displaystyle\sum_{p=1}^{\ell} z_i(C_i^p).
\end{align*}
For the remaining indices $j$, we set $U_i^j:=\emptyset$.
By the definition of $U_i^j$, we have the following lemma.
\begin{lemma}
\label{lem:num_tij_ex}
For each $i\in [k]$ and $\ell\in [d_i]$, the number of $j\in [bM]$ such that $U_i^j=C_i^\ell$ is $z_i(C_i^\ell)$.
\end{lemma}

To show that the integrality gap of \textsc{$(\mathrm{LP}_w)$} is at most $k/2$, we will construct a $k/2$-approximate integer solution to \textsc{$(\mathrm{LP}_w)$}.
Such a solution can be  constructed by taking the largest $2/k$ of the sets $U_i^{(q-1)M+1},U_i^{(q-1)M+2},\ldots,U_i^{qM}$ for each $i\in [k]$ and $q\in [b]$, and partitioning them into groups of $kb$ sets such that for each group, the sum of the characteristic vectors  is at least $w$.
This construction is based on the $k/2$-approximation algorithm for $k$-\textsc{HypVC} by Lov\'{a}sz \cite{lovasz1975}.
For the construction, we prepare the following two lemmas.

\begin{lemma}
\label{lem:zf_ex}
We have
\begin{align*}
\sum_{i\in [k],j\in [bM]}f_i(U_{i}^j)=\sum_{S\subseteq N,i\in [k]}z_i(S)f_i(S).
\end{align*}
\end{lemma}
\begin{proof}
Since $z_i(C_i^\ell)$ is the number of $j\in [bM]$ with $U_i^j=C_i^\ell$ by Lemma \ref{lem:num_tij_ex}, we have
\begin{align*}
&\sum_{S\subseteq N,i\in [k]}z_i(S)f_i(S)=\sum_{S\in \mathcal{L}_i,i\in [k]}z_i(S)f_i(S)=\sum_{\ell\in [d_i],i\in [k]}z_i(C_i^\ell)f_i(C_i^\ell)\\&=\sum_{\ell\in [d_i],i\in [k]}\left|\left\{j\in [bM]\ \middle|\  U_i^j=C_i^\ell\right\}\right|\cdot f_i(C_i^\ell)\\&=\sum_{\substack{i\in [k],j\in [bM]:\\U_i^j=C_i^\ell\ \mathrm{for\ some\ }\ell\in [d_i]}}f_i(U_i^j)=\sum_{i\in [k],j\in [bM]}f_i(U_{i}^j).
\end{align*}
The last equality follows since if $U_i^j\ne C_i^\ell$ for all $\ell=1,2,\ldots,d_i$, then we have $U_i^j=\emptyset$ by the definition.
\end{proof}

\begin{lemma}
\label{lem:condition_covering_ex}
Let $a_1,a_2,\ldots,a_k\in [M]$ be positive integers such that $\sum_{i=1}^ka_i\le M+k-1$.
Then, we have $\sum_{i=1}^k\sum_{j=1}^b \chi_{U_i^{a_{i}+(j-1)M}}\ge w$.
\end{lemma}
\begin{proof}
Suppose to the contrary that 
\begin{align}
\label{ineq:ijkb}
\left|\left\{(i,j)\in [k]\times [b]\mid e\in U_i^{a_{i}+(j-1)M}\right\}\right|< w(e)
\end{align}
for some $e\in N$.
For each $i\in [k]$, let $\gamma_i$ denote the number of $j\in [b]$ such that $e\in U_i^{a_i+(j-1)M}$.
Then, by (\ref{ineq:ijkb}) we have $\sum_{i=1}^k \gamma_i<w(e)$.
Since $U_i^1\supseteq U_i^2\supseteq \cdots \supseteq U_i^{bM}$ for each $i\in [k]$, by the definition of $\gamma_i$ it follows that $e\notin U_i^{a_i+\gamma_i M}$ for every $i\in [k]$.
Hence, we have
\begin{align}
\label{ineq:lower_mw}
&\left|\left\{(i,j)\in [k]\times [bM]\mid e\in U_i^{j}\right\}\right|\le \sum_{i=1}^k (a_i+\gamma_i M-1)\le M+k-1+(w(e)-1)\cdot M-k\notag\\&=M\cdot w(e)-1.
\end{align}
By Lemma \ref{lem:num_tij_ex}, we also have
\begin{align*}
\left|\left\{(i,j)\in [k]\times [bM]\mid e\in U_i^{j}\right\}\right|=\sum_{(i,\ell)\in [k]\times [d_i]:e\in C_i^\ell}z_i(C_i^\ell)=\sum_{(i,\ell)\in [k]\times [d_i]:e\in C_i^\ell}M\cdot y^*_i(C_i^\ell)=M\cdot w(e).
\end{align*}
This contradicts the inequality (\ref{ineq:lower_mw}).
\end{proof}

Since $M$ is divisible by $k(k-1)$, we can denote $M=k(k-1)\cdot m$ for some integer $m$.
Let $L:=\frac{2M}{k}=2m(k-1)$.
Lov\'{a}sz~\cite{lovasz1975} showed that there exists a $k\times L$ integer matrix such that integers in each row are distinct, and each column satisfy the condition in Lemma \ref{lem:condition_covering_ex}.
We provide a constructive proof of this result, since the construction will be used in the proof of Theorem \ref{thm:approx_ex}.

\begin{lemma}[Lov\'{a}sz~\cite{lovasz1975}]
\label{lem:lovasz}
There exists a $k\times L$ matrix $A=(a_{i,j})_{i=1,\ldots,k,j=1,\ldots,L}$ satisfying the following conditions:
\begin{enumerate}
    \item \label{cond1}every row of $A$ is a permutation of $\{1,2,\ldots,L\}$.
    \item \label{cond2}the sum of every column is at most $M+k-1$.
\end{enumerate}
\end{lemma}
\begin{proof}
We separately consider the two cases when $k$ is even, and when $k$ is odd.
We first consider the case when $k$ is even.
Since $k$ is even, we can denote $k=2p$ for some positive integer $p$.
For each $i\in [p]$, we define $(a_{2i-1,j})_{j=1,\ldots,L}$ as any permutation of $\{1,2,\ldots,L\}$, and define $a_{2i,j}:=L+1-a_{2i-1,j}$ for every $j\in [L]$.
Then, $(a_{i,j})_{j=1,\ldots,L}$ is a permutation of $\{1,2,\ldots,L\}$ for every $i\in [k]$, and satisfies
\begin{align*}
\sum_{i=1}^k a_{i,j}=\sum_{i=1}^p (a_{2i-1,j}+a_{2i,j})=p(L+1)=\frac{k}{2}\cdot (2m(k-1)+1)=M+\frac{k}{2}\le M+k-1
\end{align*}
for every $j\in [L]$, as required.

Consider the case when $k$ is odd.
Since $k\ge 2$ and $k$ is odd, we can denote $k=2p+1$ for some positive integer $p$.
For each integer $a\in \mathbb{Z}$, define $\mathrm{mod}(a)$ to be the unique integer $r\in [L]$ such that $a\equiv r\pmod {L}$.
For $i=1,2,3$, we define $(a_{i,j})_{j=1,\ldots,L}$ as follows:
\begin{align*}
a_{i,j}:=\begin{cases}
   j & \mathrm{if\ }i=1,\\
   \mathrm{mod}\left(\frac{L}{2}+j\right) & \mathrm{if\ }i=2, \\
   \mathrm{mod}(L-2j+2)-\left\lfloor \dfrac{2j-2}{L}\right\rfloor & \mathrm{if\ }i=3.
   \end{cases}
\end{align*}
Similarly to the case when $k$ is even, we define the remaining part as follows: for each $i=2,\ldots,p$, define $(a_{2i,j})_{j=1,\ldots,L}$ as any permutation of $\{1,2,\ldots,L\}$, and define $a_{2i+1,j}:=L+1-a_{2i,j}$ for every $j\in [L]$.
Then, it is clear that $(a_{i,j})_{j=1,\ldots,L}$ is a permutation of $\{1,2,\ldots,L\}$ for $i=1,2$ and $i\ge 4$.
For $i=3$, we have $a_{3,j}=L-2j+2$ for every $1\le j\le \frac{L}{2}$, and $a_{3,j}=2L-2j+1$ for every $\frac{L}{2}+1\le j\le L$.
Hence, $(a_{3,j})_{j=1,\ldots,L}$ is a permutation of $\{1,2,\ldots,L\}$. 
For every $1\le j\le \frac{L}{2}$, we have
\begin{align*}
&\sum_{i=1}^k a_{i,j}=\sum_{i=1}^3 a_{i,j}+\sum_{i=2}^p (a_{2i,j}+a_{2i+1,j})=j+\frac{L}{2}+j+L-2j+2+(p-1)(L+1)\\&=\left(p+\frac{1}{2}\right)L+p+1=M+\frac{k+1}{2}\le M+k-1.
\end{align*}
For every $\frac{L}{2}+1\le j\le L$, we have
\begin{align*}
&\sum_{i=1}^k a_{i,j}=\sum_{i=1}^3 a_{i,j}+\sum_{i=2}^p (a_{2i,j}+a_{2i+1,j})=j+j-\frac{L}{2}+2L-2j+1+(p-1)(L+1)\\&=\left(p+\frac{1}{2}\right)L+p=M+\frac{k-1}{2}\le M+k-1.
\end{align*}
Hence, $(a_{i,j})_{i=1,\ldots,k,j=1,\ldots,L}$ satisfies the required conditions.
\end{proof}

Let $A=(a_{i,j})_{i=1,\ldots,k,j=1,\ldots,L}$ be a $k\times L$ matrix satisfying the conditions in Lemma~\ref{lem:lovasz}.
For simplicity, we denote $[p]_0:=[p]\cup \{0\}$ for every $p\in \mathbb{Z}_{>0}$, and $T_{i,j,q}:=U_i^{a_{i,j}+qM}$ for every $i\in [k],j\in [L],$ and $q\in [b-1]_0$.
Then, for each $j\in [L]$, since we have $\sum_
{i=1}^k a_{i,j}\le M+k-1$, Lemma \ref{lem:condition_covering_ex} implies
\begin{align*}
\sum_{i=1}^{k}\sum_{q=0}^{b-1}\chi_{T_{i,j,q}}\ge w.
\end{align*}
We now show that for some $j\in [L]$, the collection of sets $\{T_{i,j,q}\}_{i=1,\ldots,k,q=0,\ldots,b-1}$ induces a $k/2$-approximate integer solution to \textsc{$(\mathrm{LP}_w)$}, which completes the proof of Theorem~\ref{thm:ig_most_lattice}.
\begin{proof}[Proof of Theorem \ref{thm:ig_most_lattice}]
By Lemma \ref{lem:zf_ex}, we have
\begin{align}
\label{ineq:zf}
&\sum_{S\subseteq N,i\in [k]}z_i(S)f_i(S)=\sum_{i\in [k],j\in [bM]}f_i(U_i^j)\ge \sum_{\substack{i\in [k],j\in [L]\\q\in [b-1]_0}}f_i\left(T_{i,j,q}\right)\notag\\&\ge L\cdot \min_{j\in [L]}\sum_{i=1}^k\sum_{q=0}^{b-1} f_i\left(T_{i,j,q}\right). 
\end{align}
Take
\begin{align*}
j^*\in \underset{j\in [L]} {\operatorname{argmin}} \sum_{i=1}^k\sum_{q=0}^{b-1} f_i\left(T_{i,j,q}\right).
\end{align*}
Then, since $z_i=M\cdot y^*_i=\frac{kL}{2}\cdot y^*_i$ for each $i\in [k]$, by (\ref{ineq:zf}) we have
\begin{align}
\label{ineq:jstar}
\frac{k}{2}\cdot \sum_{S\subseteq N,i\in [k]}y_i^*(S)f_i(S)\ge \sum_{i=1}^k\sum_{q=0}^{b-1} f_i\left(T_{i,j^*,q}\right). 
\end{align}
Since we have
\begin{align*}
\sum_{i=1}^k\sum_{q=0}^{b-1} \chi_{T_{i,j^*,q}}\ge w,
\end{align*}
there exist a collection of sets $\{S_{i,q}\}_{i=1,\ldots,k,q=0,\ldots,b-1}$ such that $S_{i,q}\subseteq T_{i,j^*,q}$ for every $i=1,\ldots,k$ and $q=0,\ldots,b-1$, and
\begin{align}
\label{eq_siq}
\sum_{i=1}^k\sum_{q=0}^{b-1} \chi_{S_{i,q}}= w.
\end{align}
By (\ref{ineq:jstar}) and the monotonicity of $f_i$, we also have
\begin{align}
\label{ineq_siq}
\frac{k}{2}\cdot \sum_{S\subseteq N,i\in [k]}y_i^*(S)f_i(S)\ge \sum_{i=1}^k\sum_{q=0}^{b-1} f_i\left(T_{i,j^*,q}\right)\ge \sum_{i=1}^k\sum_{q=0}^{b-1} f_i\left(S_{i,q}\right).
\end{align}
For each $i\in [k]$ and $S\subseteq N$, let  $y_i(S):=|\{q\in [b-1]_0\mid S=S_{i,q}\}|$.
Then, by (\ref{eq_siq}) and (\ref{ineq_siq}), $(y_1,\ldots,y_k)$ is an integer  feasible solution to \textsc{$(\mathrm{LP}_w)$} satisfying
\begin{align*}
\frac{k}{2}\cdot \sum_{S\subseteq N,i\in [k]}y_i^*(S)f_i(S)\ge \sum_{i=1}^k\sum_{q=0}^{b-1} f_i\left(S_{i,q}\right)=\sum_{S\subseteq N,i\in [k]} y_i(S)f_i(S).
\end{align*}
Thus, the integrality gap of \textsc{$(\mathrm{LP}_w)$} is at most $k/2$.
\end{proof}

The proof of Theorem \ref{thm:ig_most_lattice} provides the construction of a $k/2$-approximate integer solution to \textsc{$(\mathrm{LP}_w)$}.
The construction works as follows.
We first compute a rational optimal solution $(y^*_1,\ldots,y^*_k)$ to \textsc{$(\mathrm{LP}_w)$} whose supports are chains, and define $U_i^j$ for each $i\in [k]$ and $j\in [bM]$.
We then construct a matrix $A=(a_{i,j})_{i=1,\ldots,k,j=1,\ldots,L}$ as in the proof of Lemma \ref{lem:lovasz}, and compute an index $j^*$ minimizing $\sum_{i=1}^k\sum_{q=0}^{b-1} f_i\left(T_{i,j,q}\right)$ in the proof of Theorem~\ref{thm:ig_most_lattice}.
Finally, we obtain a collection of sets $\{S_{i,q}\}_{i=1,\ldots,k,q=0,\ldots,b-1}$, which induces a $k/2$-approximate integer solution $(y_1,\ldots,y_k)$ to \textsc{$(\mathrm{LP}_w)$}.

We prove Theorem \ref{thm:approx_ex} by showing that the above construction can be done in polynomial time.
\begin{proof}[Proof of Theorem \ref{thm:approx_ex}]
We can obtain a rational optimal solution $(y_1^*,\ldots,y_k^*)$ to \textsc{$(\mathrm{LP}_w)$} whose supports $\mathcal{L}_1,\ldots,\mathcal{L}_k$ are chains in polynomial time via the ellipsoid method \cite{chekuri2011icalp} (see Appendix \ref{app:convex_program} for details).
Then, we define $U_i^j$ for every $i\in [k]$ and $j\in [bM]$ from the members of the chains $\mathcal{L}_1,\ldots,\mathcal{L}_k$, and define a matrix $A=(a_{i,j})_{i=1,\ldots,k,j=1,\ldots,L}$ as in the proof of Lemma \ref{lem:lovasz}.
If we can compute
\begin{align*}
j^*\in \underset{j\in [L]} {\operatorname{argmin}} \sum_{i=1}^k\sum_{q=0}^{b-1} f_i\left(T_{i,j,q}\right)
\end{align*}
in polynomial time, then we can find in polynomial time an integer feasible solution $(y_1,\ldots,y_k)$ to \textsc{$(\mathrm{LP}_w)$} such that
\begin{align*}
\frac{k}{2}\cdot \sum_{S\subseteq N,i\in [k]}y_i^*(S)f_i(S)\ge \sum_{S\subseteq N,i\in [k]} y_i(S)f_i(S).
\end{align*}
Indeed, we can construct such $(y_1,\ldots,y_k)$ from $T_{i,j^*,q}$ for $i\in [k]$ and $q\in [b-1]_0$ as follows.
As in the proof of Theorem \ref{thm:ig_most_lattice}, there exists a collection of sets $\{S_{i,q}\}_{i=1,\ldots,k,q=0,\ldots,b-1}$ such that $S_{i,q}\subseteq T_{i,j^*,q}$  for every $i\in [k]$ and $q\in [b-1]_0$, $S_{i,q}\supseteq S_{i,q+1}$ for every $i\in [k]$ and $q\in [b-2]_0$, and
\begin{align}
\label{eq:xsw}
\sum_{i=1}^k\sum_{q=0}^{b-1} \chi_{S_{i,q}}= w.
\end{align}
Such a collection $\{S_{i,q}\}_{i=1,\ldots,k,q=0,\ldots,b-1}$ can be obtained in polynomial time as follows: for each $e\in N$ such that 
\begin{align*}
|\{(i,q)\in [k]\times [b-1]_0\mid e\in T_{i,j^*,q}\}|> w(e),
\end{align*} we repeatedly remove $e$ from $T_{i,j^*,q}$ containing $e$ one by one in decreasing order of $q$ until the strict inequality does not hold.
Then, we finally obtain a collection $\{S_{i,q}\}_{i=1,\ldots,k,q=0,\ldots,b-1}$ satisfying (\ref{eq:xsw}).
As in the proof of Theorem \ref{thm:ig_most_lattice}, for each $i\in [k]$, define $y_i\in \mathbb{Z}^{2^N}_{\ge 0}$ by 
\begin{align*}
y_i(S):=|\{q\in [b-1]_0\mid S=S_{i,q}\}|
\end{align*}
for each $S\subseteq N$.
Since $S_{i,q}\supseteq S_{i,q+1}$ for every $i\in [k]$ and $q\in [b-2]_0$, such $y_i$ can be obtained in polynomial time from $\{S_{i,q}\}_{q=0,\ldots,b-1}$ for every $i\in [k]$.
Moreover, as in the proof of Theorem \ref{thm:ig_most_lattice}, $(y_1,\ldots,y_k)$ is a $k/2$-approximate integer feasible solution to \textsc{$(\mathrm{LP}_w)$}.
Therefore, it suffices to show that $j^*$ can be computed in polynomial time.

When $k$ is even, by the construction of the matrix $A$ in the proof of Lemma \ref{lem:lovasz}, each pair of rows $(a_{2i-1,j})_{j=1,\ldots,L}$ and $(a_{2i,j})_{j=1,\ldots,L}$ can be defined independently of the other pairs.
Similarly, when $k$ is odd, the triple of rows $(a_{1,j})_{j=1,\ldots,L}$, $(a_{2,j})_{j=1,\ldots,L}$, and $(a_{3,j})_{j=1,\ldots,L}$, as well as each pair of rows $(a_{2i,j})_{j=1,\ldots,L}$ and $(a_{2i+1,j})_{j=1,\ldots,L}$ can be defined independently of the other pairs and the triple.
Hence, to find $j^*$,  it suffices to compute $j\in [L]$ minimizing the sum of $\sum_{q=0}^{b-1}f_i(T_{i,j,q})$ over the rows $i$ in each pair or triple.
For fixed $i\in [k]$, since $|\mathcal{L}_i|\le n+1$, the number of distinct $U_i^j$ is at most $n+1$.
Since $U_i^1\supseteq U_i^2\supseteq \cdots \supseteq U_i^{bM}$, this implies that the number of $j\in [L-1]$ such that $U_i^{j+qM}\ne U_i^{j+1+qM}$ for some $q\in [b-1]_0$ is at most $n$.
Hence, for each $i\in [k]$, the number of distinct collections of sets $(T_{i,j,q})_{q=0,\ldots,b-1}$ is at most $n+1$. 
Therefore, for each pair or triple of rows $i$, there are at most $O(n^3)$ possible combinations of $(T_{i,j,q})_{q=0,\ldots,b-1}$.
Hence, we can find $j\in [L]$ minimizing the sum of $\sum_{q=0}^{b-1}f_i(T_{i,j,q})$ over each pair or triple of $i$ in polynomial time by enumerating $O(n^3)$ such combinations that are realizable for a common $j\in [L]$, and comparing the values of the sums.
Thus, we can find $j^*$ in polynomial time.
\end{proof}

\section{Applications and hardness}
\label{sec:app}
In this section, we provide two applications of Theorems \ref{thm:ig_most_lattice} and \ref{thm:approx_ex}, and prove Theorems \ref{thm:ig_atleast} and \ref{thm:approx_hardness}.
In Section \ref{subsec:k-poly}, we provide an upper bound on the integrality gap of the dual linear program of weighted $k$-polymatroid intersection.
In Section \ref{subsec:vc}, we prove Theorems \ref{thm:ig_atleast} and \ref{thm:approx_hardness}.
The proofs are based on an approximation preserving reduction from $k$-\textsc{HypVC}, with the aid of the result in Section \ref{subsec:k-poly}.
In Section \ref{subsec:bcover}, we provide a $k/2$-approximation algorithm for the minimum-weight $b$-vertex cover problem in a $k$-partite hypergraph.

\subsection{Dual-LP of weighted $k$-polymatroid intersection}
\label{subsec:k-poly}
In this section, we show that Theorems \ref{thm:ig_most_lattice} and \ref{thm:approx_ex} can be applied to the dual linear program of weighted $k$-polymatroid intersection.
Let $N$ be a finite set and $k\ge 2$ a positive integer.
For each $i\in [k]$, let $f_i:2^{N}\to \mathbb{R}_{\ge 0}$ be a monotone submodular function with $f_i(\emptyset)=0$.
Let $w\in \mathbb{Z}^N$ be a weight vector.
The \textit{weighted $k$-polymatroid intersection problem} is defined as follows:
\begin{align*}
(\mathrm{LP}_{k\text{-poly}})\ \ \ \ \ \mathrm{maximize}&\ \ \ w^\top x\notag\\
\mathrm{subject\ to}&\ \ \ x(S)\le f_i(S)\ \mathrm{for\ every\ }S\subseteq N\ \mathrm{and\ }i\in [k],\notag\\&
\ \ \ x\ge 0,\notag
\end{align*}
where $x(S):=\sum_{e\in S}x(e)$ for every $S\subseteq N$.
When $k=2$, the linear program $(\mathrm{LP}_{k\text{-poly}})$ is exactly the weighted polymatroid intersection problem (see, e.g., \cite{fujishige2005,schrijver2003}).
We consider the following dual linear program of $(\mathrm{LP}_{k\text{-poly}})$:
\begin{align*}
(\mathrm{DLP}_{k\text{-poly}})\ \ \ \ \ \mathrm{minimize}&\ \ \ \sum_{S\subseteq N,i\in [k]}y_i(S) f_i(S)\\
\mathrm{subject\ to}&\ \ \ y_i\in \mathbb{R}^{2^N}_{\ge 0}\ \mathrm{for\ every\ }i\in [k],\\&\ \ \ \sum_{S\subseteq N,i\in [k]}y_i(S) \chi_S\ge w.
\end{align*}

When $k=2$, it is well-known that the system of linear constraints defining  $(\mathrm{LP}_{k\text{-poly}})$ is totally dual integral, that is, $(\mathrm{DLP}_{k\text{-poly}})$ has an integer optimal solution for any $w\in \mathbb{Z}^N$ (see, e.g., \cite{schrijver2003}).
The following result extends this to general $k$:
\begin{theorem}
\label{thm:dual_kpoly}
The integrality gap of $(\mathrm{DLP}_{k\text{-poly}})$ is at most $k/2$.
Moreover, there exists a polynomial-time algorithm to find an integer feasible solution to $(\mathrm{DLP}_{k\text{-poly}})$ whose objective value is at most $k/2$ times the optimum.
\end{theorem}
\begin{proof}
For any feasible solution $(y_1,\ldots,y_k)$ to $(\mathrm{DLP}_{k\text{-poly}})$, since $y_i\ge 0$ for every $i\in [k]$, we have
\begin{align}
\label{ineq:y0}
\sum_{S\subseteq N,i\in [k]}y_i(S)\chi_S\ge 0.
\end{align}
Define $w^+\in \mathbb{Z}^N_{\ge 0}$ by $w^+(e):=\max\{w(e),0\}$ for every $e\in N$.
Then, by (\ref{ineq:y0}), replacing $w$ by $w^+$ in $(\mathrm{DLP}_{k\text{-poly}})$ does not change the feasible region.
Hence, without loss of generality, we may assume that $w\ge 0$.
By Theorems \ref{thm:ig_most_lattice} and \ref{thm:approx_ex}, it suffices to show that $(\mathrm{DLP}_{k\text{-poly}})$ has an optimal solution and an optimal integer solution that are also feasible for \textsc{$(\mathrm{LP}_w)$}.
Take an optimal (or optimal integer) solution $y=(y_1,\ldots,y_k)$ to $(\mathrm{DLP}_{k\text{-poly}})$.
We show that $y$ can be converted to another optimal (or optimal integer) solution $y^*=(y^*_1,\ldots,y^*_k)$ that is feasible for \textsc{$(\mathrm{LP}_w)$}.
From now, we only discuss the case when $y$ is an optimal integer solution, because the same argument also works for the case when $y$ is an optimal solution.
For each $e\in N$, define
\begin{align*}
\lambda(e):=\sum_{S\subseteq N,i\in [k]: e\in S}y_i(S).
\end{align*}
If $\lambda(e)=w(e)$ for all $e\in N$, then we are done.
Suppose that for some $e'\in N$, we have $\lambda(e')>w(e')$.
Take $\delta_{i,S}\in [0,y_i(S)]$ for every $i\in [k]$ and $S\subseteq N$ with $e'\in S$, satisfying
\begin{align*}
\sum_{S\subseteq N,i\in [k]: e'\in S}\delta_{i,S}=\lambda(e')-w(e').
\end{align*}
Since we have
\begin{align*}
0<\lambda(e')-w(e')\le \lambda(e')=\sum_{S\subseteq N,i\in [k]: e'\in S}y_i(S),
\end{align*}
such a collection of $\delta_{i,S}$ exists.
Moreover, since $\lambda(e')$ and $w(e')$ are integers, such $\delta_{i,S}$ can be chosen to be integers.
Then, decrease $y_{i}(S)$ by $\delta_{i,S}$ and increase $y_{i}(S\setminus \{e'\})$ by $\delta_{i,S}$ for every $i\in [k]$ and $S\subseteq N$ with $e'\in S$. 
Let $y'=(y'_1,\ldots,y'_k)$ be the solution to $(\mathrm{DLP}_{k\text{-poly}})$ obtained by this way.
Then, we have $y'_i(S)\ge 0$ for every $i\in [k]$ and $S\subseteq N$, and
\begin{align*}
\sum_{S\subseteq N,i\in [k]:e\in S}y'_i(S)=\sum_{S\subseteq N,i\in [k]:e\in S}y_i(S)\ge w(e)
\end{align*}
for every $e\in N\setminus \{e'\}$, and
\begin{align*}
\sum_{S\subseteq N,i\in [k]:e'\in S}y'_i(S)=\sum_{S\subseteq N,i\in [k]:e'\in S}y_i(S)-\sum_{S\subseteq N,i\in [k]:e'\in S}\delta_{i,S}=\lambda(e')-\lambda(e')+w(e')=w(e').
\end{align*}
Hence, $y'$ is an integer feasible solution to $(\mathrm{DLP}_{k\text{-poly}})$.
Moreover, by the monotonicity of $f_i$, we have
\begin{align*}
&\sum_{S\subseteq N,i\in [k]}y_i(S)f_i(S)-\sum_{S\subseteq N,i\in [k]}y'_i(S)f_i(S)=\sum_{S\subseteq N,i\in [k]:e'\in S}\delta_{i,S}f_i(S)-\sum_{S\subseteq N,i\in [k]:e'\in S}\delta_{i,S}f_i(S\setminus \{e'\})\\&=\sum_{S\subseteq N,i\in [k]:e'\in S}\delta_{i,S}(f_i(S)-f_i(S\setminus \{e'\}))\ge 0.
\end{align*}
Therefore, $y'$ is an optimal integer solution to $(\mathrm{DLP}_{k\text{-poly}})$ such that
\begin{align*}
\sum_{S\subseteq N,i\in [k]:e\in S}y_i'(S)=\sum_{S\subseteq N,i\in [k]:e\in S}y_i(S)
\end{align*}
for every $e\in N\setminus \{e'\}$, and
\begin{align*}
\sum_{S\subseteq N,i\in [k]:e'\in S}y_i'(S)=w(e')<\lambda(e')=\sum_{S\subseteq N,i\in [k]:e'\in S}y_i(S).
\end{align*}
By repeating this procedure for all $e\in N$ satisfying $\lambda(e)>w(e)$, we finally obtain an optimal integer solution $(y^*_1,\ldots,y^*_k)$ such that
\begin{align*}
\sum_{S\subseteq N,i\in [k]}y_i^*(S)\chi_S=w.
\end{align*}
This completes the proof.
\end{proof}

\subsection{Hardness reduction from vertex cover}
\label{subsec:vc}
In this section, we prove Theorems \ref{thm:ig_atleast} and \ref{thm:approx_hardness}.
Our proofs are based on an approximation preserving reduction from $k$-\textsc{HypVC}.
We first show that the integrality gap of $(\mathrm{DLP}_{k\text{-poly}})$ is lower-bounded by that of a natural LP-relaxation of $k$-\textsc{HypVC}, denoted by  $(\textsc{LP}_{\text{VC}})$.
Combined with the proof of Theorem \ref{thm:dual_kpoly}, this implies that the integrality gap of \textsc{$\mathrm{(LP)}$} is lower-bounded by that of $(\textsc{LP}_{\text{VC}})$, which leads to the lower bound on the integrality gap in Theorem \ref{thm:ig_atleast}.
We also give an approximation preserving reduction from $k$-\textsc{HypVC} to \textsc{Monotone-MSCA}.
Combined with the approximation hardness result for $k$-\textsc{HypVC} by Guruswami et al. \cite{guruswami2015}, this implies Theorem \ref{thm:approx_hardness}.

Let $k\ge 2$ be a positive integer and $H=(V,E)$ a $k$-partite hypergraph with a given $k$-partition $\{V_1,\ldots,V_k\}$ of $V$, that is, every edge $e\in E$ satisfies $|e\cap V_i|\le 1$ for all $i\in [k]$.
Here, $e\ne \emptyset$ for every $e\in E$.
Recall that $k$-\textsc{HypVC} asks to find a vertex cover of the minimum size, where a subset $X\subseteq V$ is called a \textit{vertex cover} if $e\cap X\ne \emptyset$ for every $e\in E$.
A natural LP-relaxation $(\textsc{LP}_{\text{VC}})$ of $k$-\textsc{HypVC} is defined as follows:
\begin{align*}
(\textsc{LP}_{\text{VC}})\ \ \ \ \ \mathrm{minimize}&\ \ \ \mathbf{1}^{\top}y\\
\mathrm{subject\ to}&\ \ \ y\in \mathbb{R}^{V}_{\ge 0},\\&\ \ \ \sum_{v\in e}y(v)\ge 1\ \mathrm{for\ every\ }e\in E.
\end{align*}
We show that $(\textsc{LP}_{\text{VC}})$ can be represented as a special case of $(\mathrm{DLP}_{k\text{-poly}})$ by setting  $N:=E,w:=\mathbf{1},$ and appropriately defining $f_1,\ldots,f_k$.
To this end, we define such $f_1,\ldots,f_k$.
For each $S\subseteq E$ such that $e\cap V_i\ne \emptyset$ for every $e\in S$, we denote $V_i(S):=\bigcup_{e\in S}e\cap V_i$.
For each $i\in [k]$, define $f_i:2^E\to \mathbb{R}_{\ge 0}$ by
\begin{align}
\label{eq:fi_def}
    f_i(S)=\begin{cases}
        |V_i(S)| & \mathrm{if\ }e\cap V_i\ne\emptyset\ \mathrm{for\ every\ }e\in S,\\
        |V|+1 & \mathrm{if\ }e\cap V_i=\emptyset\ \mathrm{for\ some\ }e\in S.
    \end{cases}
\end{align}
Intuitively, $f_i(S)$ is the number of vertices in $V_i$ needed to cover $S\subseteq E$.
We first show that $f_i$ is monotone submodular for every $i\in [k]$.
\begin{lemma}
\label{lem:fi_monosub}
$f_1,\ldots, f_k$ defined in (\ref{eq:fi_def}) are monotone submodular.
\end{lemma}
\begin{proof}
For each $i\in [k]$, it is clear that $f_i$ is monotone, that is, $f_i(X)\le f_i(Y)$ for every $X\subseteq Y\subseteq E$.
To prove the submodularity of $f_i$, take any $X,Y\subseteq E$.
If $f_i(X)=|V|+1$, then we have
\begin{align*}
f_i(X)+f_i(Y)=|V|+1+f_i(Y)\ge f_i(X\cup Y)+f_i(X\cap Y).
\end{align*}
The last inequality follows from $f_i(X\cup Y)\le |V|+1$ and the monotonicity of $f_i$.
Hence, the submodular inequality holds in this case.
If $f_i(Y)=|V|+1$, the submodular inequality holds by the same argument.
Consider the case when $f_i(X)\ne |V|+1$ and $f_i(Y)\ne |V|+1$.
Then, we have
\begin{align*}
&f_i(X)+f_i(Y)=|V_i(X)|+|V_i(Y)|=|V_i(X)\cup V_i(Y)|+|V_i(X)\cap V_i(Y)|\\&\ge |V_i(X\cup Y)|+|V_i(X\cap Y)|=f_i(X\cup Y)+f_i(X\cap Y).
\end{align*}
Hence, $f_i$ is submodular for every $i\in [k]$.
\end{proof}

Let $N(v):=\{e\in E\mid v\in e\}$ for every $v\in V$.
For each $i\in [k]$, let
\begin{align*}
\mathcal{C}_i:=\{N(v)\mid v\in V_i\}.
\end{align*}
We are now ready to show the following key lemma:
\begin{lemma}
\label{lem:dlp_reduction}
For the linear program $(\mathrm{DLP}_{k\text{-poly}})$ with $N:=E$ and $f_1,\ldots,f_k$ defined as in (\ref{eq:fi_def}), there exists an optimal solution $(\tilde{y}_1,\ldots,\tilde{y}_k)$ such that its support $\tilde{\mathcal{L}}_i:=\{S\subseteq E\mid \tilde{y}_i(S)>0\}$ is included in $\mathcal{C}_i$ for every $i\in [k]$.
Moreover, there exists an optimal integer solution $(\hat{y}_1,\ldots,\hat{y}_k)$ to $(\mathrm{DLP}_{k\text{-poly}})$ such that the support of $\hat{y}_i$ is included in $\mathcal{C}_i$ for every $i\in [k]$.
\end{lemma} 
\begin{proof}
Let $y=(y_1,\ldots,y_k)$ be a feasible solution to $(\mathrm{DLP}_{k\text{-poly}})$ with $N:=E$ and $f_1,\ldots,f_k$ defined as in (\ref{eq:fi_def}).
It suffices to show that $y$ can be converted into another feasible solution $y^*=(y^*_1,\ldots,y^*_k)$ such that the support of $y^*_i$ is included in $\mathcal{C}_i$ for every $i\in [k]$, and the objective value of $y^*$ is at most that of $y$, and the conversion preserves the integrality if $y$ is integer. 
From now, we only consider the case when $y$ is integer, because the same argument also works for the case when $y$ is non-integer.

Let $\mathcal{L}_i:=\{S\subseteq E\mid y_i(S)>0\}$ for every $i\in [k]$.
We define 
\begin{align*}
\mathcal{T}_i:=\{S\subseteq E\mid e\cap V_i\ne \emptyset \mathrm{\ for\ every\ }e\in S\}
\end{align*}
for every $i\in [k]$.
We first convert $y$ to another integer feasible solution $y'=(y_1',\ldots,y_k')$ such that the support of $y_i'$ is included in $\mathcal{T}_i$ for every $i\in [k]$, and the objective value of $y'$ is at most that of $y$.
If $\mathcal{L}_i\subseteq \mathcal{T}_i$ for every $i\in [k]$, then we are done.
Suppose that there exists $X\in \mathcal{L}_j\setminus \mathcal{T}_j$ for some $j\in [k]$.
Since $e\ne \emptyset$ for every $e\in E$, $X$ can be partitioned into $X_1,X_2,\ldots,X_k$ such that $e\cap V_i\ne \emptyset$ for every $e\in X_i$ and $i\in [k]$.
Let $\alpha:=y_j(X)$.
Then, increase $y_i(X_i)$ by $\alpha$ for every $i\in [k]$, and decrease $y_j(X)$ by $\alpha$.
Let $z=(z_1,\ldots,z_k)$ denote the solution obtained by this way.
Then, $z$ is an integer feasible solution to $(\mathrm{DLP}_{k\text{-poly}})$.
We also have
\begin{align*}
&\sum_{S\subseteq E,i\in [k]}y_i(S)f_i(S)-\sum_{S\subseteq E,i\in [k]}z_i(S)f_i(S)=\alpha f_j(X)-\alpha\sum_{i=1}^k f_i(X_i)\\&\ge \alpha(|V|+1)-\alpha\sum_{i=1}^k |V_i|=\alpha>0.
\end{align*}
Hence, $z$ is an integer feasible solution to $(\mathrm{DLP}_{k\text{-poly}})$ such that the objective value of $z$ is at most that of $y$, and $X$ is not included in the support of $z_j$.
By repeating this procedure for all such $X\in \mathcal{L}_i\setminus \mathcal{T}_i$ and $i\in [k]$, we finally obtain an integer feasible solution $y'=(y_1',\ldots,y_k')$ such that its support $\mathcal{L}_i':=\{S\subseteq E\mid y_i'(S)>0\}$ is included in $\mathcal{T}_i$ for every $i\in [k]$, and the objective value of $y'$ is at most that of $y$.

We next convert $y'$ to another integer feasible solution $y^*=(y_1^*,\ldots,y_k^*)$ to $(\mathrm{DLP}_{k\text{-poly}})$ such that the support of $y_i^*$ is contained in $\mathcal{C}_i$ for every $i\in [k]$, and the objective value of $y^*$ is equal to that of $y'$.
If $\mathcal{L}'_i\subseteq \mathcal{C}_i$ for every $i\in [k]$, then we are done.
Otherwise, there exists $Y\in \mathcal{L}_j'\setminus \mathcal{C}_j$ for some $j\in [k]$.
Let $\beta:=y_j'(Y)$.
Then, increase $y'_j(N(v))$ by $\beta$ for every $v\in V_j(Y)$, and decrease $y_j'(Y)$ by $\beta$.
Let $z'=(z_1',\ldots,z'_k)$ denote the solution obtained by this way.
Then, we have
\begin{align*}
\sum_{S\subseteq E,i\in [k]}z'_i(S)\chi_S-\sum_{S\subseteq E,i\in [k]}y_i'(S)\chi_S=\sum_{v\in V_j(Y)}\beta\chi_{N(v)}-\beta\chi_{Y}\ge 0.
\end{align*}
Hence, $z'$ is an integer feasible solution to $(\mathrm{DLP}_{k\text{-poly}})$.
We also have
\begin{align*}
&\sum_{S\subseteq E,i\in [k]}z'_i(S)f_i(S)-\sum_{S\subseteq E,i\in [k]}y'_i(S)f_i(S)=\sum_{v\in V_j(Y)}\beta f_j(N(v))-\beta f_j(Y)\\&= \sum_{v\in V_j(Y)}\beta-\beta \cdot |V_j(Y)|=0.
\end{align*}
Hence, $z'$ is an integer feasible solution to $(\mathrm{DLP}_{k\text{-poly}})$ such that the objective value of $z'$ is equal to that of $y'$, and $Y$ is not included in the support of $z_j'$.
By repeating this procedure for all such $Y\in \mathcal{L}_i'\setminus \mathcal{C}_i$ and $i\in [k]$, we finally obtain an integer feasible solution $y^*=(y_1^*,\ldots,y_k^*)$ such that the support of $y^*_i$ is included in $\mathcal{C}_i$ for every $i\in [k]$, and the objective value of $y^*$ is equal to that of $y'$, that is, at most that of $y$.
Hence, $y^*$ is a desired integer feasible solution.
\end{proof}

Consider the linear program $(\mathrm{DLP}_{k\text{-poly}})$ with $N:=E, w:=\mathbf{1}$, and $f_1,\ldots,f_k$ defined as in (\ref{eq:fi_def}).
Then, by Lemma \ref{lem:dlp_reduction}, $(\mathrm{DLP}_{k\text{-poly}})$ has an optimal solution and an optimal integer solution such that the $i$-th supports of the solutions are included in $\mathcal{C}_i$ for every $i\in [k]$.
Hence, $(\mathrm{DLP}_{k\text{-poly}})$ can be reduced to the following linear program restricted to $\mathcal{C}_i$ for $i\in [k]$:
\begin{align*}
(\mathrm{DLP}_{\mathcal{C}})\ \ \ \ \ \mathrm{minimize}&\ \ \ \sum_{S\in \mathcal{C}_i,i\in [k]}y_i(S) f_i(S)\\
\mathrm{subject\ to}&\ \ \ y_i\in \mathbb{R}^{\mathcal{C}_i}_{\ge 0}\ \mathrm{for\ every\ }i\in [k],\\&\ \ \ \sum_{S\in \mathcal{C}_i,i\in [k]}y_i(S) \chi_S\ge \mathbf{1}.
\end{align*}
By the definition of $\mathcal{C}_i$ and $f_i$ for $i\in [k]$, $(\mathrm{DLP}_{\mathcal{C}})$ can also be represented as follows:
\begin{align*}
(\mathrm{DLP}_{\mathcal{C}})\ \ \ \ \ \mathrm{minimize}&\ \ \ \sum_{v\in V_i,i\in [k]}y_i(N(v))\\
\mathrm{subject\ to}&\ \ \ y_i\in \mathbb{R}^{\mathcal{C}_i}_{\ge 0}\ \mathrm{for\ every\ }i\in [k],\\&\ \ \ \sum_{v\in V_i,i\in [k]}y_i(N(v)) \chi_{N(v)}\ge \mathbf{1}.
\end{align*}
Then, $(\mathrm{DLP}_{\mathcal{C}})$ is equivalent to $(\textsc{LP}_{\text{VC}})$.
Indeed, for a solution $y=(y_1,\ldots,y_k)$ to $(\mathrm{DLP}_{\mathcal{C}})$, define $y'\in \mathbb{R}_{\ge 0}^V$ by $y'(v):=y_i(N(v))$ for every $v\in V_i$ and $i\in [k]$.
Then, the objective function and the constraints  can be written as 
\begin{align*}
\sum_{v\in V_i,i\in [k]}y_i(N(v))=\sum_{v\in V}y'(v),
\end{align*}
and
\begin{align*}
\sum_{v\in V_i,i\in [k]:v\in e}y_i(N(v))=\sum_{v\in e}y'(v)\ge 1
\end{align*}
for every $e\in E$.
The above objective function and constraints are exactly same as those for $(\textsc{LP}_{\text{VC}})$.
Hence, $(\mathrm{DLP}_{\mathcal{C}})$ is equivalent to $(\textsc{LP}_{\text{VC}})$.
This implies that the integrality gap of $(\mathrm{DLP}_{\mathcal{C}})$ is equal to that of $(\textsc{LP}_{\text{VC}})$.
By the result of Aharoni et al. \cite{aharoni1996} that the integrality gap of $(\textsc{LP}_{\text{VC}})$ is $k/2-o(1)$, the integrality gap of $(\mathrm{DLP}_{\mathcal{C}})$ is also $k/2-o(1)$.
Since the integrality gap of $(\mathrm{DLP}_{k\text{-poly}})$ for general case is lower-bounded by that of $(\mathrm{DLP}_{\mathcal{C}})$, we obtain the following result:
\begin{theorem}
\label{thm:dlp_lowerb}
The integrality gap of $(\mathrm{DLP}_{k\text{-poly}})$ is at least $k/2-o(1)$ even when $w=\mathbf{1}$.
\end{theorem}

Note that  the integrality gap lower bound in Theorem \ref{thm:dlp_lowerb} matches the upper bound in Theorem \ref{thm:dual_kpoly}.
By the proof of Theorem \ref{thm:dual_kpoly}, $(\mathrm{DLP}_{k\text{-poly}})$ has an optimal solution and an optimal integer solution that are also optimal to \textsc{$(\mathrm{LP}_w)$}.
Hence, Theorem \ref{thm:dlp_lowerb} implies the following result, which includes Theorem \ref{thm:ig_atleast}:
\begin{corollary}
\label{cor:lpw_lowerb}
The integrality gap of \textsc{$(\mathrm{LP}_w)$} is at least $k/2-o(1)$.
In particular, the integrality gap of \textsc{$\mathrm{(LP)}$} is at least $k/2-o(1)$.
\end{corollary}
By Corollary \ref{cor:lpw_lowerb}, the upper bounds on the integrality gaps of \textsc{$\mathrm{(LP)}$} and \textsc{$(\mathrm{LP}_w)$} in Theorems \ref{theorem:igmost} and \ref{thm:ig_most_lattice} are tight.

We next prove Theorem \ref{thm:approx_hardness} via an approximation preserving reduction from $k$-\textsc{HypVC}.
Guruswami et al. \cite{guruswami2015} showed that assuming the Unique Games Conjecture (UGC), $k$-\textsc{HypVC} is NP-hard to approximate within a factor of $k/2-\epsilon$ for any $\epsilon >0$ and fixed $k\ge 2$.
Hence, since $(\mathrm{DLP}_{\mathcal{C}})$ is equivalent to $(\textsc{LP}_{\text{VC}})$, $(\mathrm{DLP}_{\mathcal{C}})$ restricted to integer solutions is also NP-hard to approximate within a factor of $k/2-\epsilon$ for any $\epsilon >0$ and fixed $k\ge 2$ under UGC.

Consider the linear program $(\mathrm{DLP}_{k\text{-poly}})$ with $N:=E, w:=\mathbf{1}$, and $f_1,\ldots,f_k$ defined as in (\ref{eq:fi_def}).
Recall that $(\mathrm{DLP}_{k\text{-poly}})$ defined above has an optimal integer solution such that the $i$-th support of the solution is included in $\mathcal{C}_i$ for every $i\in [k]$.
In addition, by the proof of Lemma \ref{lem:dlp_reduction}, an integer feasible solution to $(\mathrm{DLP}_{k\text{-poly}})$ can be converted to another integer feasible solution satisfying the above support condition without increasing the objective value.
Hence, there exists an approximation preserving reduction from $(\mathrm{DLP}_{\mathcal{C}})$ restricted to integer solutions to $(\mathrm{DLP}_{k\text{-poly}})$ restricted to integer solutions. 
Therefore, by the approximation hardness result of $(\mathrm{DLP}_{\mathcal{C}})$ restricted to integer solutions, we obtain the following result:
\begin{theorem}
\label{thm:kpoly_hard}
Assuming the Unique Games Conjecture, it is NP-hard to compute an integer feasible solution to $(\mathrm{DLP}_{k\text{-poly}})$ whose objective value is at most $k/2-\epsilon$ times the objective value of an optimal integer solution for any $\epsilon >0$ and fixed $k\ge 2$, even when $w=\mathbf{1}$.
\end{theorem}

Note that Theorem \ref{thm:kpoly_hard} implies that the approximation ratio in Theorem \ref{thm:dual_kpoly} is optimal under UGC.
By the proof of Theorem \ref{thm:dual_kpoly}, $(\mathrm{DLP}_{k\text{-poly}})$ has an optimal integer solution that is also optimal for \textsc{$(\mathrm{LP}_w)$}.
Since any integer feasible solution to \textsc{$(\mathrm{LP}_w)$} is also feasible to $(\mathrm{DLP}_{k\text{-poly}})$, Theorem \ref{thm:kpoly_hard} implies the following result, which includes Theorem \ref{thm:approx_hardness}:
\begin{corollary}
\label{cor:msca_hardness}
Assuming the Unique Games Conjecture, it is NP-hard to compute an integer feasible solution to \textsc{$(\mathrm{LP}_w)$} whose objective value is at most $k/2-\epsilon$ times the objective value of an optimal integer solution for any $\epsilon >0$ and fixed $k\ge 2$.
In particular, assuming the Unique Games Conjecture, \textsc{Monotone-MSCA} is NP-hard to approximate within a factor of $k/2-\epsilon$ for any $\epsilon >0$ and fixed $k\ge 2$.
\end{corollary}
\label{thm:hardness}
By Corollary \ref{cor:msca_hardness}, assuming UGC, the approximation ratios in Theorems \ref{thm:approx} and \ref{thm:approx_ex} are optimal.

\subsection{Minimum-weight $b$-vertex cover problem in $k$-partite hypergraphs}
\label{subsec:bcover}
In this section, we provide a $k/2$-approximation algorithm for the minimum-weight $b$-vertex cover problem in a $k$-partite hypergraph.
Lov\'{a}sz \cite{lovasz1975} gave a $k/2$-approximation algorithm for the unweighted vertex cover problem in a $k$-partite hypergraph.
Daboul et al. \cite{daboul2023} extended Lov\'{a}sz's algorithm to the minimum-weight vertex cover problem, and obtained a $k/2$-approximation.
Our result is a generalization of these approximability results.

Let $k\ge 2$ be a positive integer and $H=(V,E)$ a $k$-partite hypergraph with a given $k$-partition $\{V_1,\ldots,V_k\}$ of $V$ such that $e\ne \emptyset$ for every $e\in E$.
Let $c\in \mathbb{R}^V_{\ge 0}$ be a weight vector and $b\in \mathbb{Z}^E_{\ge 0}$ a demand vector.
The \textit{minimum-weight $b$-vertex cover problem} asks to find a vector $y\in \mathbb{Z}^V_{\ge 0}$ minimizing $c^{\top}y$ such that $\sum_{v\in e}y(v)\ge b(e)$ for every $e\in E$.
For convenience, we denote this problem as $(k,b,c)$-\textsc{HypVC}.
A natural LP-relaxation of $(k,b,c)$-\textsc{HypVC} is as follows:
\begin{align*}
(\textsc{LP}_{b\text{-VC}})\ \ \ \ \ \mathrm{minimize}&\ \ \ c^{\top}y\\
\mathrm{subject\ to}&\ \ \ y\in \mathbb{R}^{V}_{\ge 0},\\&\ \ \ \sum_{v\in e}y(v)\ge b(e)\ \mathrm{for\ every\ }e\in E.
\end{align*}

As $(\textsc{LP}_{\text{VC}})$ can be represented as a special case of $(\mathrm{DLP}_{k\text{-poly}})$ in Section \ref{subsec:vc}, we show that $(\textsc{LP}_{b\text{-VC}})$ can also be represented as a special case by setting  $N:=E,w:=b,$ and appropriately defining $f_1,\ldots,f_k$.
Such $f_1,\ldots,f_k$ are defined similarly to those in Section \ref{subsec:vc} as follows.
For each $i\in [k]$, define $f_i:2^E\to \mathbb{R}_{\ge 0}$ by
\begin{align}
\label{eq:cfi_def}
    f_i(S)=\begin{cases}
        c\left(V_i(S)\right) & \mathrm{if\ }e\cap V_i\ne\emptyset\ \mathrm{for\ every\ }e\in S,\\
        c(V)+1 & \mathrm{if\ }e\cap V_i=\emptyset\ \mathrm{for\ some\ }e\in S,
    \end{cases}
\end{align}
where $c(T):=\sum_{v\in T}c(v)$ for every $T\subseteq V$.
The monotonicity and submodularity of $f_i$ easily follows from the proof of Lemma \ref{lem:fi_monosub}.
\begin{lemma}
\label{lem:cfi_monosub}
$f_1,\ldots, f_k$ defined in (\ref{eq:cfi_def}) are monotone submodular.
\end{lemma}

Consider the linear program $(\mathrm{DLP}_{k\text{-poly}})$ with $N:=E,w:=b$, and $f_1,\ldots f_k$ defined as in (\ref{eq:cfi_def}).
Then, by the proof of Lemma \ref{lem:dlp_reduction}, $(\mathrm{DLP}_{k\text{-poly}})$ has an optimal integer solution such that the $i$-th support of the solution is included in $\mathcal{C}_i$ for every $i\in [k]$, which corresponds to an optimal integer solution to $(\textsc{LP}_{b\text{-VC}})$.
In addition, by the proof of Lemma \ref{lem:dlp_reduction}, any integer feasible solution to $(\mathrm{DLP}_{k\text{-poly}})$ can be converted to another integer feasible solution satisfying the above support condition without increasing the objective value.
Hence, as in Section \ref{subsec:vc}, there is an approximation preserving reduction from $(\textsc{LP}_{b\text{-VC}})$ restricted to integer solutions to $(\mathrm{DLP}_{k\text{-poly}})$ restricted to integer solutions.
Therefore, we obtain the following result by Theorem \ref{thm:dual_kpoly}:
\begin{theorem}
\label{thm:bcover}
There exists a $k/2$-approximation algorithm for $(k,b,c)$-\textsc{HypVC}.
\end{theorem}

Assuming the Unique Games Conjecture, since $k$-\textsc{HypVC} is NP-hard to approximate within a factor of $k/2-\epsilon$ for any $\epsilon >0$ and fixed $k\ge 2$ \cite{guruswami2015}, the approximation ratio in Theorem \ref{thm:bcover} is optimal.

\section*{Acknowledgments}
This work was partially supported by JSPS KAKENHI Grant Number JP26K21170 and JST ERATO Grant Number JPMJER2301, Japan.

\section*{AI Declarations}
The author used generative AIs (specifically ChatGPT) to assist in refining the proofs, searching for references, and language editing.
All mathematical claims, proofs, and references were verified by the author.
The author takes full responsibility for the content of the submitted manuscript.

\bibliography{main}
\bibliographystyle{plain}

\appendix

\section{Convex program based on the Lov\'{a}sz extension}
\label{app:convex_program}
In this section, we show that \textsc{(LP)} is equivalent to the convex program relaxation of \textsc{MSCA} introduced by Chekuri and Ene \cite{chekuri2011icalp}.
Let $N:=\{1,\ldots,n\}$.
Let $f_i:2^N\to \mathbb{R}$ be a submodular function with $f_i(\emptyset)=0$ for each $i\in [k]$.
Recall that \textsc{(LP)} is defined as follows:
\begin{align*}
\textsc{(LP)}\ \ \ \ \ \mathrm{minimize}&\ \ \ \sum_{S\subseteq N,i\in [k]}y_i(S) f_i(S)\\
\mathrm{subject\ to}&\ \ \ y_i\in \mathbb{R}^{2^N}_{\ge 0}\ \mathrm{for\ every\ }i\in [k],\\&\ \ \ \sum_{S\subseteq N,i\in [k]}y_i(S) \chi_S=\mathbf{1}.
\end{align*}
For a set function $f:2^N\to \mathbb{R}$ with $f(\emptyset)=0$, the Lov\'{a}sz extension $\hat{f}:[0,1]^n\to \mathbb{R}$ of $f$ is defined as follows.
Take any $x\in [0,1]^n$.
Relabel the elements in $N$ so that $x_1\ge x_2\ge \cdots \ge x_n$.
For convenience, let $x_{n+1}:=0$.
Let $S_i:=\{1,2,\ldots,i\}$ for each $i\in [n]$.
The value of $\hat{f}(x)$ is defined as follows:
\begin{align*}
\hat{f}(x):=\sum_{i=1}^{n}(x_i-x_{i+1})f(S_i).
\end{align*}
Chekuri and Ene \cite{chekuri2011icalp} introduced the following convex program relaxation of \textsc{MSCA} based on the Lov\'{a}sz extension:
\begin{align*}
(\textsc{LE-Rel})\ \ \ \ \ \mathrm{minimize}&\ \ \ \sum_{i=1}^k \hat{f}_i(x^i)\\
\mathrm{subject\ to}&\ \ \ x^i\in [0,1]^n\ \mathrm{for\ every\ }i\in [k],\\&\ \ \ \sum_{i=1}^k x^i_j=1\ \mathrm{for\ every\ }j\in [n].
\end{align*}
We now show that any feasible solution $({x^1},\ldots,x^k)$ to \textsc{(LE-Rel)} can be translated into a feasible solution $(y_1,\ldots,y_k)$ to \textsc{(LP)} that has the same objective value and whose supports are chains.
Take any feasible solution $({x^1},\ldots,x^k)$ to \textsc{(LE-Rel)}.
We will describe a translation from $x^i$ to $y_i$ for each $i\in [k]$.
We first relabel the elements in $N$ so that $x^i_1\ge x^i_2\ge \cdots \ge x^i_n$.
Let $x^i_{n+1}:=0$ and $S_j:=\{1,2,\ldots,j\}$ for each $j\in [n]$.
Then, for all $S\subseteq N$ define $y_i(S)$ as follows:
\begin{align*}
y_i(S):=\begin{cases}
     x_j^i-x_{j+1}^i& \mathrm{if\ }S=S_j\ \mathrm{and\ }x_j^i-x_{j+1}^i>0\ \mathrm{for\ some\ }j\in [n],\\
    0 & \mathrm{otherwise}.
\end{cases}
\end{align*}
Then, $(y_1,\ldots,y_k)$ is a feasible solution to \textsc{(LP)} because for each $i\in [k]$ we have
\begin{align*}
\sum_{S\subseteq N}y_i(S)\chi_S=\sum_{j\in [n]}(x^i_j-x^i_{j+1})\chi_{S_j}=x^i
\end{align*}
and
\begin{align*}
\sum_{S\subseteq N,i\in [k]}y_i(S)\chi_S=\sum_{i\in [k]}x^i=\mathbf{1}.
\end{align*}
For each $i\in [k]$, we also have
\begin{align*}
\sum_{S\subseteq N}y_i(S)f_i(S)=\sum_{j\in [n]}(x_j^i-x_{j+1}^i)f_i(S_j)=\hat{f_i}(x^i).
\end{align*}
Hence, we obtain
\begin{align*}
\sum_{S\subseteq N,i\in [k]}y_i(S)f_i(S)=\sum_{i=1}^k\hat{f_i}(x^i),
\end{align*}
which implies that $(y_1,\ldots,y_k)$ has the same objective value as $(x^1,\ldots,x^k)$.
Similarly, we can translate any feasible solution to \textsc{(LP)} whose supports are chains into a feasible solution to \textsc{(LE-Rel)} that has the same objective value.

Since \textsc{(LE-Rel)} can be solved in polynomial time via the ellipsoid method \cite{chekuri2011icalp}, an optimal solution to \textsc{(LP)} whose supports are chains can be obtained in polynomial time.
Chekuri and Ene \cite{chekuri2011arxiv} provided another equivalent linear programming formulation of \textsc{MSCA} called \textsc{LE-Rel}-Primal. 
We note that \textsc{(LP)} is also equivalent to \textsc{LE-Rel}-Primal.

\end{document}